# Frequency stabilization based on $H_{13}C_{14}N$ absorption in lithium niobate micro-disk laser

Zhen Yi [a,b], Zhihao Zhang [b,*], Jianglin Guan [a,b], Guanghui Zhao [c,d], Renhong Gao [b], Botao Fu [b], Jintian Lin [c,d], Jinming Chen [b], Jian Liu [a,b], Yijie Pan [e], Ya Cheng [a,b,f,g,h,i,*]

[a] State Key Laboratory of Precision Spectroscopy, School of Physics and Electronic Science, East China Normal University, Shanghai 200062, China
[b] The Extreme Optoelectromechanics Laboratory (XXL), School of Physics and Electronic Science, East China Normal University, Shanghai 200241, China
[c] State Key Laboratory of High Field Laser Physics and CAS Center for Excellence in Ultra-Intense Laser Science, Shanghai Institute of Optics and Fine Mechanics (SIOM), Chinese Academy of Sciences (CAS), Shanghai 201800, China
[d] Center of Materials Science and Optoelectronics Engineering, University of Chinese Academy of Sciences, Beijing 100049, China
[e] Center for Advanced Measurement Science, National Institute of Metrology, Beijing 100029, China
[f] Collaborative Innovation Center of Extreme Optics, Shanxi University, Taiyuan, Shanxi 030006, China
[g] Collaborative Innovation Center of Light Manipulations and Applications, Shandong Normal University, Jinan 250358, China
[h] Hefei National Laboratory, Hefei 230088, China
[i] Joint Research Center of Light Manipulation Science and Photonic Integrated Chip of East China Normal University and Shandong Normal University, East China

Corresponding authors at: School of Physics and Electronic Science, East China Normal University, Shanghai 200241, China.
E-mail addresses: zhhzhang@phy.ecnu.edu.cn (Z. Zhang), ya.cheng@siom.ac.cn (Y. Cheng).

ABSTRACT

We demonstrate an on-chip lithium niobate micro-disk laser based on hydrogen cyanide ($H_{13}C_{14}N$) gas saturation absorption method for frequency stabilization. The laser chip consists of two main components: a micro-disk laser and a combined racetrack ring cavity. By operating on the $H_{13}C_{14}N$ P12 absorption line at 1551.3 nm, the laser frequency can be precisely stabilized. The laser demonstrates remarkable stability, achieving a best stability value of $9\times10^{-9}$. Furthermore, the short-term stability, evaluated over continuous time intervals of 35 seconds, showcases exceptional performance. Additionally, the residual drift remains well below 30 MHz.

## 1. Introduction

The on-chip laser, utilizing thin-film lithium niobate (TFLN) as its foundation, offers an array of desirable features, encompassing high power, precise single wavelength emission, narrow linewidth, extensive tunability, and a compact form factor [1-11]. Consequently, it finds applicability in diverse fields, including atomic clocks, lidar systems, artificial intelligence, and quantum information processing [12-17]. With the continuous advancement of technology, there is an increasing need for lasers that exhibit enhanced frequency stability and reduced frequency noise. Consequently, it becomes crucial to develop specialized frequency stabilization techniques specifically designed for lithium niobate lasers and seamlessly integrate them into scalable and stable ultra-large-scale photonic integrated circuits. Various frequency stabilization techniques have been effectively employed in semiconductor lasers, including Pound-Drever-Hall (PDH) stabilization [18-22], saturated absorption spectroscopy of atoms and molecules [23], fiber stabilization [24], and microcavity stabilization [25-26]. Similarly, analogous techniques have been utilized to realize hybrid integrated stabilized lasers on silicon [27-28] and silicon nitride ($Si_3N_4$) [29-30] platforms. However, there have been no reports regarding the implementation of these techniques on lithium niobate platforms.

In order to achieve frequency-stabilized single-mode laser operation on a lithium niobate platform, the initial prerequisite is to generate a single-mode laser output. Multi-mode lasers are susceptible to mode hopping, which poses a significant challenge for achieving frequency stabilization. The vernier effect, achieved through resonator combinations, can effectively address this challenge and produce a reliable single-mode laser output. Subsequently, an appropriate frequency stabilization technique must be carefully chosen. Among various methods, the saturated absorption technique offers notable advantages, including a compact chamber volume, a simple structure, and high frequency stability. Furthermore, for lasers operating at different wavelengths, one can employ specific atomic or molecular absorption cells, such as iodine (663 nm) [31-33], rubidium (780 nm) [34-36], acetylene (1530 nm) [37-39], and $H_{13}C_{14}N$ (1550 nm) [39-43]. Lastly, it is essential for the laser system to possess wavelength tunability. Lithium niobate exhibits exceptional electro-optic properties, enabling the utilization of linear electro-optic effects to apply electrical tuning signals to the optical carrier wave. This capability allows for real-time and high-speed control of the laser wavelength.

  We successfully achieve absolute frequency stabilization of a lithium niobate micro-disk laser, which fabricated by the photolithography assisted chemo-mechanical etching (PLACE) technology. Our laser system comprises a micro-disk laser and a combined racetrack ring cavity, allowing for wavelength tunability through the use of electrodes on micro-disk. Leveraging the vernier effect of the racetrack ring cavity, our laser system achieves a single-mode output. Through active stabilization techniques, we successfully stabilize the laser frequency at 1551.3 nm, specifically on the $H_{13}C_{14}N$ P12 absorption line, achieving an exceptional stability level of $9\times10^{-9}$. We assessed the short-term stability, which demonstrated a duration of 35 seconds, accompanied by a residual drift below 30 MHz Furthermore, we conducted an analysis of the laser's noise characteristics throughout the short-term stabilization process.

## 2. Structure design and fabrication

We employed the PLACE technology to fabricate the complete laser device on a thin film of lithium niobate. This technology enables the production of large-scale photonic devices with minimal transmission losses on TFLN [44]. Initially, we developed an $Er^{3+}$-doped TFLN on an insulator substrate, specifically TFLNOI (700 nm Xcut TFLN / 2 μm $SiO_2$ / 500 μm $LiNbO_3$), with an erbium ion concentration of 0.5 mol. Subsequently, we utilized a femto-second laser to directly write a laser mask pattern onto the 600 nm chromium film on the surface of the lithium niobate. Following that, we transferred this pattern onto the lithium niobate film using a chemical mechanical polishing process, subsequently removing the remaining chromium film via chemical etching. We proceeded to coat the prepared laser with a 500 nm gold film and directly write the electrode pattern using a femto-second laser. Finally, we obtained the lithium niobate laser chip, as depicted in the figure, after employing the standard cleaning method.

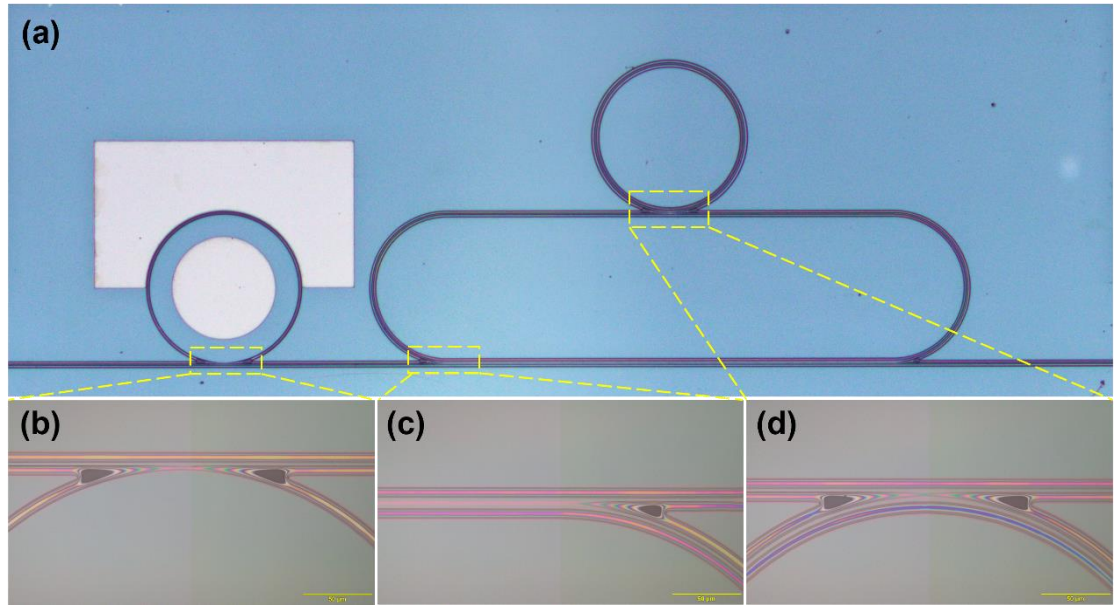

Fig. 1. (a) Optical microscope image of the TFLN micro-disk laser. (b) Zoomed view of the coupling region of the micro-disk labeled in a box of (a). (c) Zoomed view of the coupling region of the racetrack ring labeled in a box of (a). (d) Zoomed view of the coupling region of the micro-ring labeled in a box of (a).

The laser chip configuration, depicted in Figure 1(a), comprises two primary components. The initial component is a micro-disk laser responsible for laser light generation. The non-suspended micro-disk, with a radius of 200 μm, is coupled with a ridge waveguide. Figure 1(b) presents an enlarged view of the ridge waveguide, revealing its dimensions: 2.2 μm at the top and 10 μm at the bottom. A 4.5 μm coupling gap exists between the ridge waveguide and the micro-disk. The smooth surface near the coupling region results in a distinct diffraction pattern, as evident in Figure 1(b). An etched depth of approximately 300 nm is maintained to ensure appropriate restraint and efficient coupling of cavity modes. The circular electrode on the disk possesses a radius of 150 μm, while a rectangular electrode along the exterior of the micro-disk groove facilitates the application of positive and negative voltages. Laser light generated by the micro-disk propagates through a straight waveguide to the succeeding section of the structure. The second section of the configuration consists of a racetrack ring and a micro-ring, serving the purpose of frequency selection for the generated multi-mode laser. In Figure 1(c), the coupling length between the racetrack ring and the straight waveguide is measured at 1.2 mm, accompanied by a coupling gap of 4 μm. Figure 1(d) depicts a coupling gap of 4.6 μm between the racetrack ring and the micro-ring, both having a radius of 200 μm. Employing the vernier effect, the racetrack ring design incorporates two resonators that function as an interactive filter, enabling precise wavelength selection for the micro-disk laser. This innovative design effectively suppresses multi-mode laser emission caused by high-power pump lasers and ensures the production of single-frequency laser output.

## 3. Result and discussion

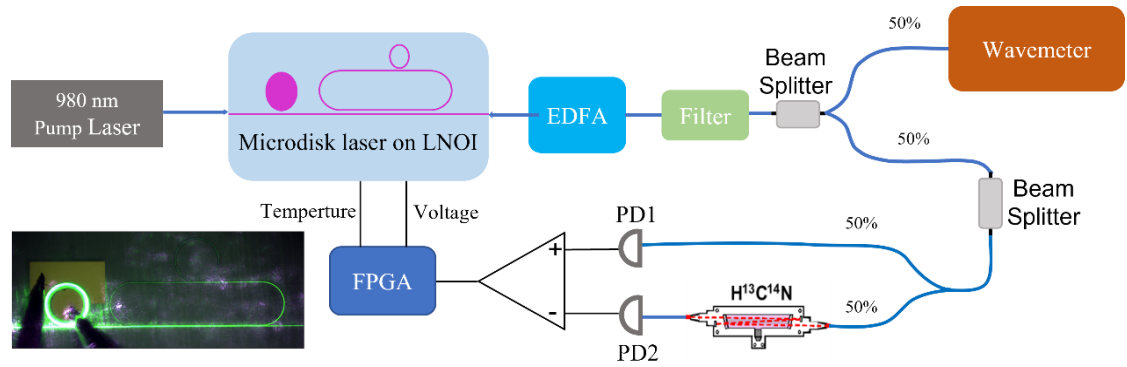

Fig. 2. Experimental setup of a delayed self-heterodyne interferometric measurement for intrinsic laser linewidth (EDFA: erbium-doped fiber amplifier; PD:photodetector). Inset: The green upconversion fluorescence of the TFLN micro-disk laser.

## 3.1 Laser frequency stabilize characterization

Figure 2 illustrates our setup, where a 980 nm laser serves as the pump light source. The pump light was effectively guided into the lithium niobate laser chip using a lens fiber. To maintain precise operating conditions, the chip was situated on a temperature control table (HC Photonics Corp. TC038-PC), guaranteeing optimal temperature stability. The generated laser light was coupled out of the chip using a lens fiber. The coupling between the lens fiber and the chip was precisely controlled using a high-precision six-axis console (Thorlabs NanoMax 601D/M). The coupled laser was amplified using an EDFA (Connet MFAS-1550-B-500-FA) and the amplified spontaneous emission (ASE) noise floor of the EDFA was effectively filtered out using an adjustable filter. This filtering process ensures that the amplified laser maintains a high-side mode rejection ratio, thus reducing unwanted noise and interference. Subsequently, the laser light was directed into a high-precision wavelength meter (Highfinesse WS-6) using a beam splitter. This setup enables the real-time measurement of the laser's frequency stability. The remaining laser light was further split into two channels using a beam splitter. One of the channels passed through the $H_{13}C_{14}N$ absorption cell and was converted into an electrical signal by the photodetector (PD2). The other channel served as a reference light and was directly converted into an electrical signal by the photodetector (PD1). The $H_{13}C_{14}N$ absorption cell has a length of 16.5 cm and operates at a pressure of 2 Torrs. The two photodetectors along with their associated circuits are integrated into a balanced photodetector (Koheron PD100B). The output signal from this balanced photodetector represents the error signal generated by the $H_{13}C_{14}N$ absorption. An FPGA processed the error signal and output a feedback voltage to an external voltage source. This feedback voltage was used to control the voltage value applied to the lithium niobate laser, thereby stabilizing the laser wavelength precisely at the apex of the absorption peak.

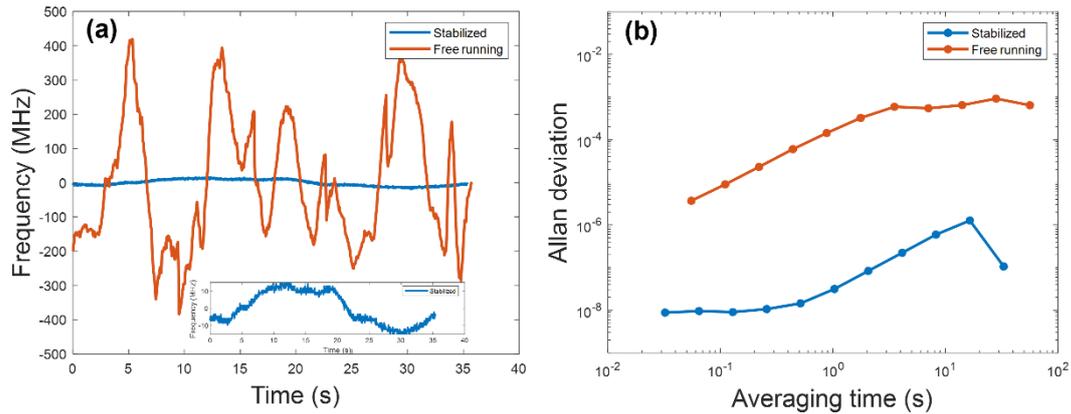

Fig. 3. (a) The recorded frequency deviations for the stabilized laser (blue) and the free running laser (orange). Inset: Zoomed view of the stabilized laser frequency deviations. (b) the overlapping Allan deviation of the frequency deviations of the stabilized and the free-running laser.

In order to evaluate the consistency of the laser's frequency, we began by fine-tuning the laser's wavelength to 1531.3 nm through the application of an initial voltage to the lithium niobate laser's electrodes. The frequency deviation of the free-running laser during the initial 35 seconds of the measurement period is depicted by the orange trace in Figure 3(a), whereas the blue trace represents the frequency deviation of the laser after stabilization. From the figure, it is evident that the frequency fluctuation range of the laser, once stabilized, decreases from 800 MHz to 30 MHz. Moreover, no pattern jumps were observed during this period. It is important to note that any occurrence of mode jumps during this time would cause a significant deviation in the laser's frequency, leading to a noticeable shift from the corresponding center frequency range depicted in the graph.

To conduct a comprehensive analysis of frequency stability across various time scales. Figure 3(b) presents the overlapping Allan bias plots of both the free-running and stabilized laser frequency deviations. As we extend the average time, it becomes evident that the frequency of the stabilized laser exhibits superior stability compared to the free-running laser. This can be attributed to the effectiveness of the stabilized scheme in detecting and correcting long-term drift. The optimal frequency stability achieved by the free-running laser is $3.7\times10^{-6}$ with an average time of 0.055 seconds. However, for stabilized laser, the frequency stability is significantly enhanced across all time scales, reaching a remarkable minimum of $9\times10^{-9}$ with an average time of 0.12 seconds.

Frequency instability exhibits a significant correlation with temperature. Prolonged injection of pump light into the laser chip results in the accumulation of thermal effects. Despite the utilization of a high-precision temperature controller to regulate chip temperature. Accurately controlling the temperature drift caused by thermal effects remains challenging. Moreover, the experimental employment of spatial coupling is more susceptible to external environmental fluctuations compared to packaged lasers. To achieve prolonged frequency stabilization, it is imperative to mitigate the influence of environmental factors and other perturbations.

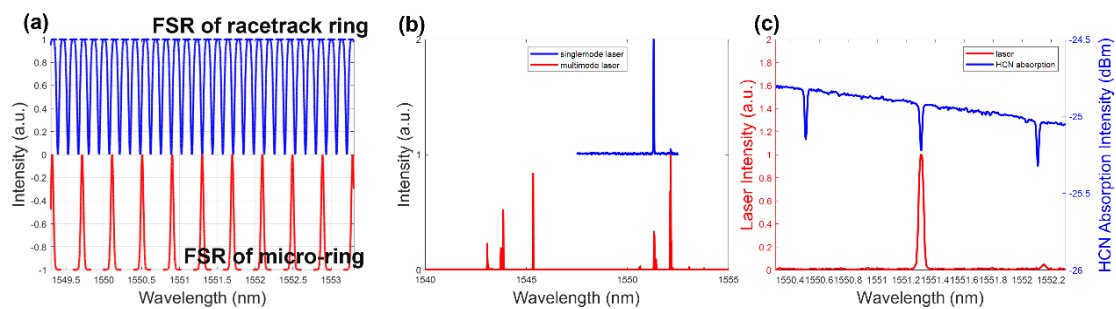

Fig. 4. (a) Schematic diagram illustrating the free spectral range (FSR) of the racetrack ring (blue) and the micro-ring (red); (b) Output spectra of the single micro-disk (red) and the LiNbO$_3$ laser chip (blue). (c) Output spectra around the emission laser at 1531.2 nm and the absorption spectra of H$_{13}$C$_{14}$N.

*3.2 Laser structure analysis*

Theoretical transmission spectra of the racetrack ring resonator (depicted by the blue line) and the micro-ring resonator (depicted by the red line) in the frequency-selective structure of the lithium niobate laser chip are depicted in Figure 4a. The free spectral range (FSR) varies between the two resonators due to their distinct perimeters. Satisfying the relation: FSR=c/n$_g$L. Where c is the speed of light, n$_g$ is the effective group index, and L is the perimeter of the resonator. Consequently, the varying transmission functions of the two resonators lead to a vernier effect, enabling the effective selection of a specific channel from the multiple lasing peaks within the micro-disk cavity. The combined cavity facilitates the frequency selection of the micro-disk laser. Subsequently, we conducted experimental verification of the frequency-selective structure within the lithium niobate laser. First, we measured the micro-disk structure within the chip, wherein a high-power (800 mW) pump light at 980 nm was injected into a singular micro-disk. We chose the high power pump light is to produce a higher power laser that can meet the transmission needs in frequency stabilization system, and limited by the performance of EDFA, lower power lasers cannot be amplified. The resulting excited spectrum (depicted by the red line) is presented in Figure 4b, illustrating a multimode mode. Subsequently, we injected the pump light, maintaining the same power, into the laser chip. The resulting excited laser is depicted by the blue line in Figure 3b, clearly indicating that the frequency selection structure within the laser chip effectively filters out the single-mode laser that corresponds to the wavelength of the $H_{13}C_{14}N$ absorption peak. We employed a pump laser with a wavelength of 980 nm, coupling the pump light into the waveguide via the input port using a lens fiber. The polarization state of the pump light was adjusted utilizing a polarization controller. The resulting laser signal is subsequently coupled through a lens fiber at the output port and directed to a spectrum analyzer (YOKOGAMA AQ6370D), possessing a spectral resolution of 0.02 nm. The laser output depicted by the red line, measured at a wavelength of 1551.3 nm, aligns with the P12 absorption peak of $H_{13}C_{14}N$, as visually depicted by the blue line in Figure 4c. The absorption curve is measured by injecting the ASE light source directly into the HCN gas chamber at 25 ℃, and the corresponding absorption intensity is 0.3 dBm, which is related to the theoretical absorption intensity in that the resolution of the spectrometer is only 0.05 nm.

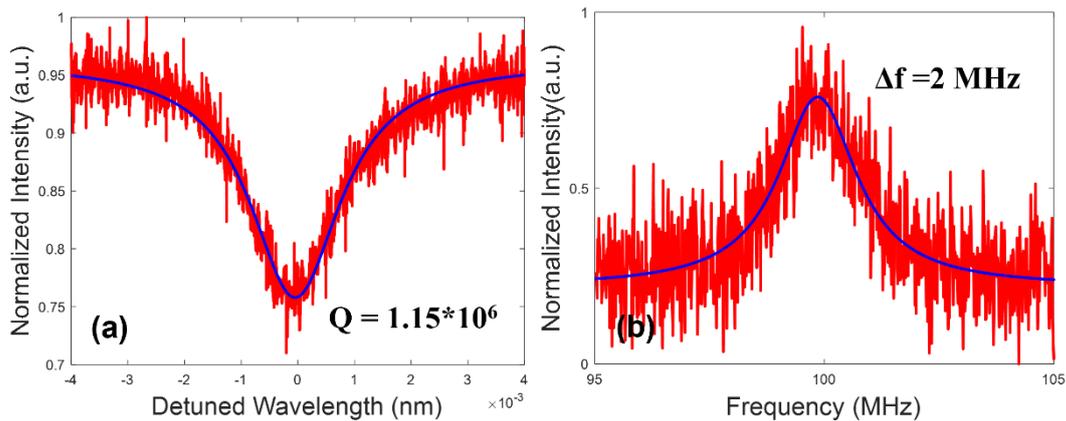

Fig. 5. (a) Lorentz fitting (blue curve) reveals a loaded Q factor of $1.15 \times 10^6$ at the wavelength of 1551.3 nm. (b) The Lorentz fitting (blue curve) of the detected beating signal featuring a laser linewidth of 2 MHz.

*3.3 Laser quality factor and linewidth characterization*

We employed the Topas 1550 tunable laser to quantify the quality factor of the entire lithium niobate laser chip. An EDFA (Connet MFAS-1550-B-500-FA) was utilized to amplify the 1551.3 nm signal light. The amplified signal was then efficiently coupled to the laser chip through a lensed fiber, and we measured the transmitted power using a reliable photodetector (Thorlabs DXM30AF). The resulting transmission spectrum, as depicted in Figure 5a, exhibited a remarkable quality factor of $1.15 \times 10^6$. Subsequently, to accurately determine the laser linewidth, we employed an optical delayed self-heterodyne interferometer configuration. This configuration involved two arms, forming an unbalanced interferometer setup. The delay arm was constructed using a single-mode fiber spanning a length of 5 km, complemented by an in-line polarization controller to ensure optimal performance. Conversely, the other arm was seamlessly connected to an acousto-optic modulator (AOM-1550-100M, Csrayzer Inc.), responsible for shifting the optical frequency by precisely 100 MHz. The laser light emitted by the lithium niobate laser was amplified by an EDFA, guided through the interferometer, and captured by a high-speed photodetector boasting a bandwidth of 1 GHz. The acquired output was then subjected to real-time spectrum analysis. Figure 5b visually presents the Lorentz fitting curve of the output signal, centered around a frequency of 100 MHz, enabling us to calculate the laser linewidth as 2 MHz/2 = 1 MHz.

## 4. Conclusions

In conclusions, we have successfully developed an on-chip lithium niobate laser that utilizes $H_{13}C_{14}N$ saturation absorption. This groundbreaking laser showcases outstanding short-term stability, boasting an unprecedented value of $9\times10^{-9}$, accompanied by a residual drift of fewer than 30 MHz. It is crucial to acknowledge that the laser's long-term stability is vulnerable to external environmental influences, thereby warranting enhancements in packaging. Moving forward, our research endeavors will encompass an in-depth exploration of various frequency bands for stabilizing lithium niobate lasers.

**Funding**

National Key R&D Program of China (2019YFA0705000), Innovation Program for Quantum Science and Technology (2021ZD0301403), National Natural Science Foundation of China (Grant Nos. 12334014,12192251), Science and Technology Commission of Shanghai Municipality (NO.21DZ1101500), Shanghai Municipal Science and Technology Major Project(2019SHZDZX01), supported by Fundamental Research Funds for the Central Universities

## CRediT authorship contribution statement

**Zhen Yi**: Writing – original draft, Data curation, Visualization, Investigation. **Zhihao Zhang**: Writing – original draft, Conceptualization, Data curation, Supervision, Resources, Validation, Methodology. **Jianglin Guan**: Data curation. **Guanghui Zhao:** Data curation. **Renhong Gao**: Data curation. **Botao Fu:** Software. **Jintian Lin**: Conceptualization, Writing – review & editing. **Jinming Chen:** Resources. **Jian Liu:** Resources. **Yijie Pan:** Writing – review & editing, Methodology. **Ya Cheng**: Writing – review & editing, Supervision, Funding acquisition, Conceptualization

## Declaration of competing interest

The authors declare that they have no known competing financial interests or personal relationships that could have appeared to influence the work reported in this paper.

## Data availability

Data will be made available on request.